\documentclass[doublecol]{epl2} 
\usepackage{graphicx}

\title{Dark matter decay and the abundance of ultracompact minihalos}

\author{Yu-Peng. Yang\inst{1,2,3} \thanks{E-mail: \email{yyp@chenwang.nju.edu.cn}}
\and Gui-Lin. Yang\inst{1,3} \thanks{E-mail: \email{yanggl@chenwang.nju.edu.cn}}
\and Hong-Shi. Zong\inst{1,3,4} \thanks{E-mail: \email{zonghs@chenwang.nju.edu.cn}}
}
\shortauthor{Y. Yang \etal}

\institute{                   
  \inst{1} Department of Physics, Nanjing University - Nanjing, 210093, China\\
  \inst{2} School of Astronomy and Space Science, Nanjing University - Nanjing, 210093, China\\
  \inst{3} Joint Center for Particle, Nuclear Physics and Cosmology - Nanjing, 210093, China\\
\inst{4} State Key Laboratory of Theoretical Physics, Institute of Theoretical Physics, CAS, Beijing 100190, China}
\pacs{98.80.Cq}{Early Universe, dark matter halos are formed in the early universe}

\abstract{
Ultracompact minihalos would be formed if there are larger density perturbations 
($0.0003 < \delta\rho/\rho < 0.3$) in the earlier epoch. The density profile of them is steeper than 
the standard dark matter halos. If the dark matter can annihilate or decay 
into the standard particles, e.g., photons, these objects would be the potential 
astrophysical sources. In order to be consistent with the observations, such as 
$Fermi$, the abundance of ultracompact minihalos must be constrained. On the 
other hand, the formation of these objects has very tight relation 
with the primordial 
curvature perturbations on smaller scale, so the fraction of ultracompact minihalos is very 
important for modern cosmology. In previous works, the studies are focused on the 
dark matter annihilation for these objects. But if the dark matter is not annihilated, 
the dark matter decay is 
another important possible case. On the other hand, the abundance of ultracompact minihalos 
is related to many other parameters, such as the mass of dark matter, 
the decay channels and the density profile of dark matter halo. 
One of the important aspects of this work is that we investigate the $\gamma$-ray 
signals from nearby ultracompact minihalos due to dark matter decay and 
another important aspect is to study in detail
how the different decay channels and density profiles affect the constraints on the abundance of ultracompact 
minihalos.}

\begin{document}

\maketitle

\section{Introduction}	
The structure formation is one of the important fields of modern cosmology. 
It is well known that the present structures of our observational 
cosmos come from the earlier density perturbations which is produced during the 
inflation. The amplitude of these density perturbations are  $\delta \rho/\rho 
\sim 10^{-5}$. If the density perturbations at earlier epoch are larger than 
0.3, the primordial black holes (PBHs) would be formed~\cite{pbh}. 
Recently, Ricotti and Gould proposed that if the density perturbations are larger than 
$10^{-3}$ \footnote{Actually, this value depends on the redshift and the scale 
of density perturbations. For more details one can see Ref.~\cite{1110.2484}.}
but smaller than 0.3, one new kind of dark matter structure 
called ultracompact minihalos (UCMHs) 
would be formed \cite{0908.0735}. The formation time of these objects is earlier and the 
density is larger than the standard dark matter halos. If the dark matter particles 
consist of weakly interacting massive particles (WIMPs), it is excepted that 
UCMHs would have notable effect on the cosmological evolution due to the dark 
matter annihilation or decay. In Refs.~\cite{yyp_prd,yyp_epjp,zhangdong}, 
the authors studied the impact of 
UCMHs on the cosmological ionization and obtained the constraints on the abundance 
of them. If high energy photons are produced by the dark matter 
annihilation, the UCMHs will become one kind of astrophysical sources and they 
can contribute to the $\gamma$-ray 
background~\cite{pat_prl,1006.4970,1110.2484,yyp_jcap,1002.3444,
1002.3445,1003.3466}. The authors of Refs.~\cite{pat_prl,1006.4970,1110.2484} 
investigated 
these effects and obtained the constraints on the fraction of UCMHs. They found 
that the strongest constraint is $\mathrm{f_{UCMHs}} \sim 4 \times 10^{-7}$ for the mass 
of UCMHs $\mathrm{M_{UCMHs}} \sim 7 \times 10^{3} \mathrm{M_{\odot}}$. In Ref.~\cite{yyp_jcap}, the authors 
investigated the contributions of UCMHs to the extragalactic gamma-ray 
background and found that the constraints increase with the dark matter mass: 
$\mathrm{f_{UCMHs}} \sim 10^{-5}$ and $\sim 10^{-3}$ 
for the dark matter mass $\mathrm{M_{\chi}} \sim 10 \mathrm{GeV}$ and 
$\sim 1 \mathrm{TeV}$, respectively. 
These constraints are stronger than the results obtained 
from the CMB data by about one order of magnitude (see the Fig.~(4) in Ref.~\cite{yyp_jcap}).
\footnote{For both cases, the constraints on the UCMHs abundance are independent of
the mass of UCMHs.}
If the dark matters are not annihilated, due to the steep density profile 
of UCMHs, they would be detected by microlensing observations \cite{lifd}. 
The formation of UCMHs has very closer relation with many aspects, such as the primordial density 
perturbations on the smaller scales 
($k = 5 \sim 10^{8} \un{Mpc^{-1}}$) which cannot be constrained by the cosmic 
microwave background (CMB), LSS and Lyman-$\alpha$ (see the Fig.~(6) of Ref.~\cite{1110.2484}). 
Moreover, the UCMHs can also be used to constraints the non-gaussianity~\cite{1211.7361}, 
so the fraction of UCMHs is very important for modern cosmology. 

As the necessary component of cosmology, dark matter has been confirmed 
by many observations. But the essence of them is still unknown and 
there are many models at present. One of the important models is that they are composed 
of WIMPs, such as neutralino, which comes from the supersymmetric 
extension of the standard model~\cite{dm_1,dm_2,dm_3}. 
According to this theory, they are stable and can annihilate into the standard particles. 
Except for these "annihilation models", there are also other models and one important of them 
is "decay models", e.g. decaying gravitinos in R-parity violating SUSY models~\cite{decay_1,decay_2}. 
These particles are unstable and can also decay into the standard particles. 
The lifetime of them is usually longer or comparable than the cosmic age. 
All of the previous works mainly focused on the impact of 
UCMHs due to the dark matter annihilation. But, as was mentioned above, if the dark matters are 
not annihilated, the decay would be another important possible case, 
and it will be considered in this work firstly for UCMHs. We use the WMAP-7 years data 
to obtain the constraints on the decay rate $\Gamma(s^{-1})$ (or the lifetime $\tau (s)$) 
and then use these results to perform the following calculations.  

Another important factor which affects the final constraints on the abundance of UCMHs 
is the density profile of Milky Way's dark matter halo. The density profile of dark matter halo 
has been "determined" by simulations and observations. One of the popular 
models is the NFW profile which is obtained through the simulations~\cite{nfw}. 
But it diverges in the center as $r \to 0$, 
and this can be avoided by other models, such as isothermal~\cite{iso} and Einasto~\cite{ein}. 
On the other hand, there are also many observations to constrain
the density profile~\cite{pro_1,pro_2,pro_3}. For the abundance of UCMHs, 
because the final constraints are related to the density profile of dark matter halo, 
in this work we will investigate how the different density profiles 
affect the fraction of UCMHs.

This paper is organized as follows.  In Sec. II we show the integrated $\gamma$-ray flux 
from UCMHs due to dark matter decay. In Sec. III
we discuss how the related parameters affect the constraints on UCMHs fraction and the conclusion is given in Sec. IV.

\section{The $\gamma$-ray signals from UCMHs due to dark matter decay}
After UCMHs are seeded during the radiation dominated epoch, the dark matter 
particles will be accreted by the radial infall. The mass of UCMHs changed 
slowly until the matter is dominated. It can be written in the form \cite{0908.0735,pat_prl}

\begin{equation}
M_\mathrm{UCMHs}(z) = M_i \left(\frac{1 + z_\mathrm{eq}}{1+z}\right),
\end{equation}
where $M_i$ is the mass within the perturbation scale
 at the time of matter-radiation equality.

\begin{figure}
\onefigure[scale=0.65]{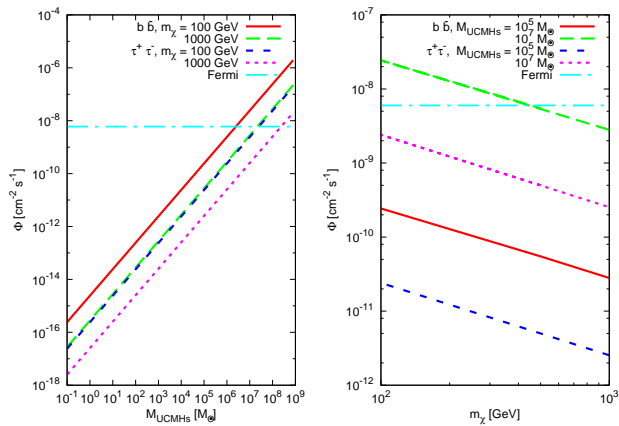}
\caption{The integrated $\gamma$-ray signals above 0.1GeV from nearby UCMH 
($d$ = 10kpc) 
due to dark matter decay. Two 
decay channels are shown: $b \bar b$ and $\tau^+ \tau^-$ and the decay rate is 
$\Gamma  = 10^{-26} s^{-1}$. The point source sensitivity of $Fermi$ are also shown.}
\label{dec_exam}
\end{figure}

The density profile can be obtained through the simulation 
\cite{profile_of_ucmhs_1,profile_of_ucmhs_2,0908.0735} 

\begin{equation}
\rho(r,z) = \frac{3f_\chi M_\mathrm{UCMHs}(z)}{16\pi R(z)^\frac{3}{4}r^\frac{9}{4}},
\end{equation}
where ${R(z)}=0.019\left(\frac{1000}{z+1}\right)\left(\frac{M(z)}
{\mathrm{M}_\odot}\right)^\frac{1}{3} \mathrm{pc}$ and 
$f_{\chi} = \frac{\Omega_{DM}}{\Omega_b+\Omega_{DM}} = 0.83$~\cite{wmap} 
is the dark matter fraction. Due to the structure 
formation effect, the mass of UCMHs will stop increasing at recent time and 
in this work we assumed the corresponding redshift is $z = 10$ \cite{pat_prl,
1110.2484}. The radius of UCMHs is $R(z=10) \sim 0.01 M_i^\frac{1}{3}$ and 
$R \sim $1 kpc for $M_i = 10^6 M_\odot$. So in the following discussions we 
will also treat the UCMHs as point sources \cite{0908.0735,pat_prl,1110.2484}. 
The density profile of UCMHs is obtained by assuming the prefect 
radial infall. But after the formation of UCMHs, it is not always the true cases 
and the angular momentum will be important. It means that this effect will 
make the density at the center smaller and a core would be 
formed instead of  $\rho \to \infty$ for $r \to 0$. 
We accept the analysis process in Ref.~\cite{1110.2484} and set the minimum radius 
$r_\mathrm{min} \sim 10^{-7}(M_\mathrm{UCMHs(z=0)}/M_\odot)^{-0.06}R_\mathrm{UCMHs(z=0)}$. 
For the smaller radius, we assume that the density is a constant 
$\rho_{r<r_\mathrm{min}} = \rho(r_\mathrm{min})$.  
The integrated flux of $\gamma$-ray signals from nearby UCMHs can 
be written as 

\begin{equation}
\label{phi}
\Phi = \frac{1}{4\pi d^2}\frac{\Gamma}{m_\chi}\sum^i \int^{m_{\chi}}_{E_{th}} 
B_{f_i} \frac{dN_i}{dE}dE\int \rho(r) d^{3}r
\end{equation}
where the summation is for all decay channels, $d$ is the distance from the earth. 
$\frac{dN}{dE}$ is the energy spectrum of dark matter decay, which can be 
obtained by the public code: DarkSUSY
\footnote{http://www.physto.se/~edsjo/darksusy/}. $B_{f_i}$ is the branching ratio 
of each decay channel. In this work, we consider each decay channel 
separately and set $B_f$ = 1. The $\mathrm{E_{th}}$ is the threshold value of the detector. 
Here we consider the $Fermi$ observation and choose the threshold value $\mathrm{E_{th}}$ = 100 MeV.   
In Fig.~\ref{dec_exam}, the integrated $\gamma$-ray signals above 0.1GeV 
from UCMH are shown. Two decay channels are plotted: $b \bar b$ and 
$\tau^+ \tau^-$. For the decay rate, we have chosen $\Gamma = 10^{-26} s^{-1}$ and 
the distance of UCMH is $d = 10 \mathrm{kpc}$. In order to compare with 
the observations, the point source sensitivity of $Fermi$ are also shown 
\footnote{ http://fermi.gsfc.nasa.gov/science/instruments
/table1-1.html}
($\Phi_{(\mathrm{E > 100 MeV})} = 4.0 \times 10^{-9} 
\mathrm{cm ^2 s^{-1}}$ for one year observation times, $5\sigma$ confidence level).
From the figure one can see that the integrated flux is proportional to the 
mass of UCMHs ($\mathrm{\Phi} \propto\mathrm{M_{UCMHs}}$) and inversely 
proportional to the dark matter mass ($\mathrm{\Phi} \propto 1/\mathrm{m_{\chi}}$). 
Actually, these characters can be seen apparently from Eq.~({\ref{phi}}). Therefore, 
for the fixed distance, the final flux would exceed the point source sensitivity 
of $Fermi$ for the larger UCMHs mass or smaller dark matter mass. On the other 
hand, it also can be seen that the final flux depends on the decay channels 
while the difference are not very large \cite{1110.2484}. 
Moreover, one should realize that the integrated flux is also related to the 
dark matter decay rate which now have been constrained by many observations
\cite{1205.5283,1009.5988,0912.4504,0906.1571,0912.0663,1205.5918,0709.2299,1004.2036,
1110.6236,1205.6474}. 

\section{Constraints on the Fraction of UCMHs}
If one single UCMH exists in the Milky Way dark matter halo, the upper limits 
on the fraction of UCMHs for the non-detection results 
(e.g. $Fermi$) can be written as~{\bf \cite{1006.4970}}, 
$f_\mathrm{UCMHs} = 
\frac{\mathrm{M_{UCMHs}}(z=0)}{\mathrm{M_{DM}}(r < d_{obs})}$, where
$d_{obs}$ is the distance that the $\gamma$-ray signals from UCMHs can be 
observed by the detector and $\mathrm{M_{DM}}(r < d_{obs})$ is the mass of Milky Way 
within this radius. On the other hand, if UCMHs indeed exist within 
the distance $d_{obs}$, but they are not observed due to 
the limitation of the detector, the upper limits on the fraction of UCMHs 
will become~\cite{1110.2484}

\begin{equation}
\label{cons}
f_\mathrm{{UCMHs}} = \frac{f_{\chi}M_\mathrm{{UCMHs(z=0)}}}{M_\mathrm{{MW}}}
\frac{\mathrm{log}(1-y/x)}{\mathrm{log}(1-M_\mathrm{{d<d_{obs}}}/M_\mathrm{{MW}})}
\end{equation}
where $y$ and $x$ are the confidence level corresponding to the 
$f_\mathrm{UCMHs}$ and the detector, respectively. $M_\mathrm{MW}$ is the 
dark matter halo mass of Milky Way. More general and slightly improved 
form is presented in Ref.~\cite{1211.7361} (eq. A2). We find that 
the corresponding changes of the results for our work can be neglected safely.
We also consider the limits from 
the galactic diffuse emission which is very important for the smaller UCMHs. 
For the very large distance of UCMHs, the constraints from the extragalactic sources 
are also considered (for more details, one can see Eqs.(28) and (29) 
in Ref.~\cite{1110.2484}). Following Ref.~\cite{0912.0663},
we use three kinds of density profile for the Milky Way, NFW, Isothermal (Iso), 
Einasto (Ein) and use the value of parameters in that reference. 
We also assume that the abundance of UCMHs is the same everywhere \cite{pat_prl,1110.2484}. 

From the discussions in Section 2, it can also be seen that several parameters can affect 
the final integrated $\gamma$-ray flux. So, the final constraints on the abundance of UCMHs 
will also be affected by these parameters. In this section, we will discuss 
in detail how these parameters affect the final constraints. One of the important 
parameters is the dark matter decay rate ($\Gamma$). 
It has no theoretically defined value 
but has to be constrained by observations. There are many ways to give the 
constraints on this parameters including using the extragalactic $\gamma$-ray 
or CMB. In this work, we use the WMAP-7 years data to obtain the constraints 
on the decay rate. For the detailed discussions on the effect of dark matter decay on the 
CMB one can see Ref.~\cite{xuelei}. Here we only give a general description. 
The dark matter can decay into the standard particles such as photons ($\gamma$), electrons ($e^-$) 
and positrons ($e^+$) and so on. These particles will have interaction with 
the content of cosmos, such as the interaction between the photons and the 
hydrogen atoms which are formed after the recombination ($z \sim 1100 $). One of the effects is to
delay the recombination. Moreover, the dark matter decay would be one kind of 
sources of reionization. Therefore, the evolution of ionization fraction including 
the dark matter decay can be written as \cite{xuelei}

\begin{equation}
\label{reion}
\frac{dx_e}{dz} = \frac{1}{(1+z)H(z)}[R_s(z)-I_s(z)-I_{DM}(z)]
\end{equation}
where $R_s$ and $I_s$ are the standard recombination rate and ionization rate. 
$I_{DM}$ is the ionization rate from the dark matter decay and it is 
related to the decay rate $I_{DM} \propto \zeta\Gamma $, 
where $\zeta$ stands for the fraction of the energy
which has been injected into the baryonic gas by the dark 
matter decay.
The effect of the dark matter decay on the evolution of ionization fraction 
will influence the power 
spectrum of CMB and we can use the WMAP data to obtain the constraints on the 
decay rate. We modified the public 
code CAMB \footnote{http://camb.info/} to include the effect of dark matter 
decay and the code COSMOMC \footnote{http://cosmologist.info/cosmomc/}
to obtain the constraints on the parameters. 
We use the seven
years WMAP data \cite{wmapdata}, and the data from ACBAR \cite{acbar}, 
Boomerang \cite{boom}, CBI \cite{cbi} and VSA \cite{vsa} experiments and 7 cosmological 
parameters, \{$\Omega_b h^2$, $\Omega_d h^2$, $\theta$, $\tau$, $n_s$, $A_s$, $\zeta\Gamma$\}, 
where $\Omega_b h^2$ and $\Omega_d h^2$  are the density of 
baryon and dark matter, $\theta$ is the ratio of
the sound horizon at recombination to its angular diameter distance 
multiplied by 100, $\tau$ is the optical depth, 
$n_s$ is the spectral index and $A_s$ is the amplitude of
the primordial density perturbation power spectrum. 
For the constraints on the dark matter decay rate, 
following the methods in Refs.~\cite{xuelei,lezhang}, we have used
a flat prior for $\zeta\Gamma(\times10^{-26} s^{-1}):[0,100.]$ and it is 
enough for our purpose.
The final results are listed in 
Tab.~\ref{ta1}. From these results, we can obtain the constraints 
on the decay rate for the WMAP-7 years: 
$\zeta \Gamma < 0.77 \times 10^{-25} s^{-1} (2 \sigma)$. Generally speaking, $\zeta$ 
depends on the redshift and different decay channels. In the following work, we set $\zeta$ = 1 and this is enough 
for our consideration. 
The dark matter decay rate can also be constrained by the gamma-ray 
or the X-ray observations. Generally speaking, these results depend on the 
decay channels and the density profile of dark matter halos. For some decay 
channels, the limits on the decay rate are 
stronger than our results. For example, for the gamma-ray observations of the 
 Fornax by the $Fermi$, the limit on the dark matter lifetime 
for the $b \bar b$ channel is $\tau \sim 4 \times 10^{26} s$ for the dark matter 
mass $m_{\chi} \sim 300$ GeV \cite{1110.6236}. All these results can be 
applied easily to our work.

\begin{largetable}
\caption{Posterior constraint on the cosmological parameters including dark matter decay 
from WMAP-7 years observation. }

\label{ta1}

\begin{tabular}{lccccccr} 
\hline
\hline
Parameters & $100\Omega_{b}h^2$ & $\Omega_{c}h^2$ & $ \theta_{S}$ & $\tau$ & $\Gamma(\times10^{-24} s^{-1})$ 
& $n_s$ & log[$10^{10}A_s$] \\
\hline
mean & 2.231 & 0.119 & 1.040 & 0.075 & 0.035 & 0.958 & 3.113 \\
2$\sigma$ lower & 2.143 & 0.113 & 1.036 & 0.042 & 0.000 & 0.937 & 3.050 \\
2$\sigma$ upper & 2.321 & 0.125 & 1.104 & 0.105 & 0.077 & 0.980 & 3.178 \\
\hline
\hline
\end{tabular}
\end{largetable}

\begin{figure*}
\begin{center}
\onefigure[scale=0.85]{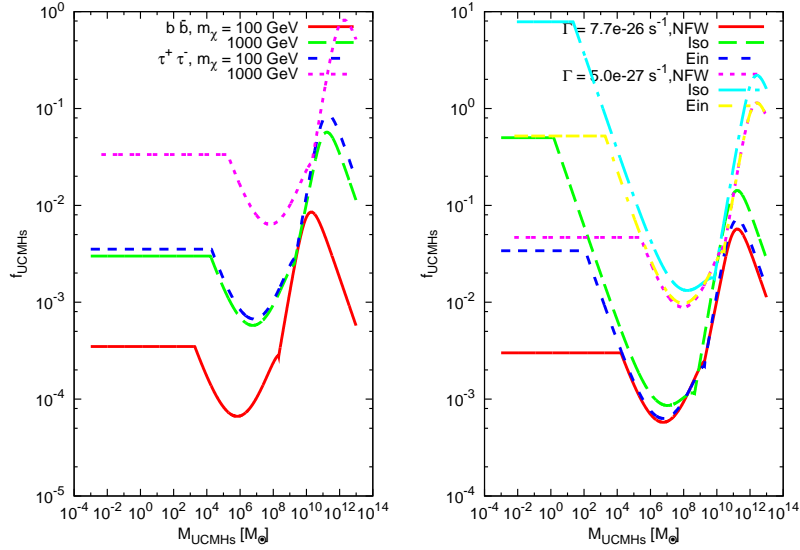}
\caption{Upper limits on the fraction of UCMHs for different dark matter 
mass, decay channels, density profile and decay rate. Left: two dark matter 
mass $m_{\chi}$ = 100 GeV, 1 TeV and decay channels $b \bar b$, $\tau^+ \tau^-$ 
for NFW density profile for our Milky Way dark matter halo are shown. For 
this plot, the dark matter decay rate is $\Gamma = 7.7 \times 10^{-26} s^{-1}$. 
Right: different dark matter decay rate $\Gamma = 7.7 \times 10^{-26} s^{-1}$, 
$5.0 \times 10^{-27} s^{-1}$ and density profile for our Milky Way 
dark matter halos NFW, Iso, Ein are shown. In this figure, we have fixed 
the decay channel $b \bar b$ and dark matter mass $m_{\chi}$ = 1 TeV.}
\label{dmcfdd}
\end{center}
\end{figure*}

In this work, we consider two typical decay channels: $b \bar b$ and $\tau^+ \tau^-$.
In order to obtain the final constraints, 
we first obtain the distance ($d_{obs}$ in Eq.~({\ref{cons}})) for the fixed UCMHs mass 
where the integrated gamma-ray flux does not exceed the point source sensitivity
of $Fermi$, and then we can use Eq.~({\ref{cons}}) to obtain the final results. 
The final constraints on the fraction of UCMHs 
for 95\% confidence level for different dark matter mass, decay channels, 
decay rate and density profiles are shown in Fig.~\ref{dmcfdd}.

From Fig. 2, it
can be seen that the strongest constraints comes from the $b \bar b$ channel 
and the dark matter mass $\mathrm{m_{\chi} = 100 GeV}$. The corresponding 
fraction is $\mathrm{f_{UCMHs}} \sim 5 \times 10^{-5}$ with the mass of 
UCMHs $\mathrm{M_{UCMHs}} \sim 10^{6} \mathrm{M_{\odot}}$. 
For a fixed mass of dark matter and UCMHs, the constraints for the 
lepton decay channels are weaker. This is due to the smaller integrated photon 
number for these decay channels. For a fixed density profile of Milky Way's dark matter 
halo and dark matter mass, the limits for the fraction are weaker for the 
smaller decay rate (longer lifetime). It can also be seen that the constraints obtained in 
this work are weaker than those of other works which considered the case of dark matter 
annihilation\cite{1006.4970,1110.2484}. 
The main reason is that the strength of dark matter decay is 
proportional to the density rather than the density squared. 
One should note that even though 
the final results are weaker than the cases of annihilation, however, 
if the dark matter are not annihilated, then the decay could be another 
very important case. So, the constraints 
obtained by us are also very significant for the dark matter decay models. 

\begin{figure*}
\begin{center}
\onefigure[scale=0.85]{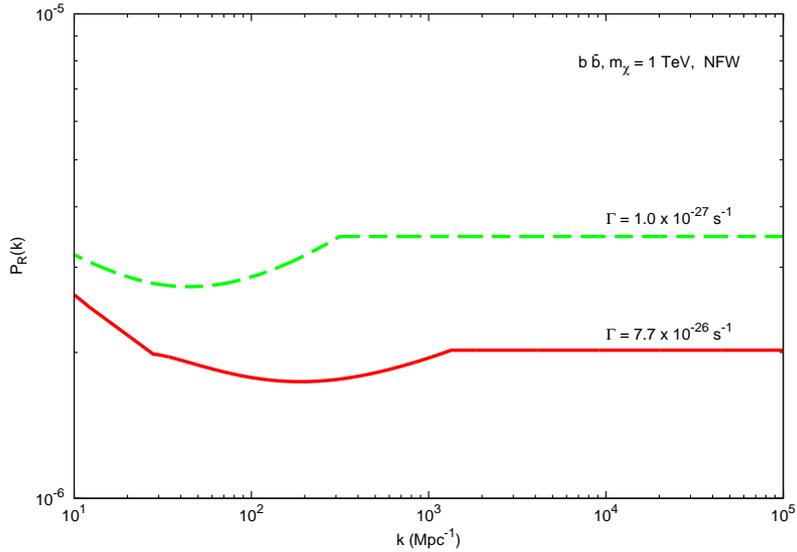}
\caption{Conservative limit on the primordial curvature perturbations obtained for 
the dark matter mass $m_{\chi}$ = 1 TeV and $b \bar b$ channel, and NFW density profile 
for our Milky Way dark matter halo. Here, we have set $\delta_\mathrm{min}$ 
= 0.006 which corresponds to the largest value obtained in ~\cite{1110.2484} 
for the formation of UCMHs at all scales.}
\label{cons_pri}
\end{center}
\end{figure*}

The constraints on the fraction of UCMHs can be translated into the 
limits on the primordial curvature perturbations on smaller scales. The process 
of calculation is given in Ref.~\cite{1110.2484}. In this work, we follow their methods and give 
the conservative limits on the primordial curvatures perturbations which are shown 
in Fig.~\ref{cons_pri}. \footnote{Detailed discussions for these constraints are prepared 
for another work.} For this plot, we set the dark matter mass 
$m_{\chi}$ = 1 TeV and choose the $b \bar b$ decay channel and NFW dark matter halo model for the Milky Way. 
For the decay rate, we have used two values, $\Gamma = 7.7 \times 10^{-26} \mathrm{s^{-1}}$ which 
has been obtained by us using the CMB data, and $\Gamma = 1.0 \times 10^{-27} \mathrm{s^{-1}}$ which 
corresponds to the allowed value obtained from other observations (e.g. gamma-ray observations \cite{1205.6474}). 
Moreover, the latter value is also within the allowed regions obtained by us. 
From this figure it can be seen that the 
strongest limit is $\mathcal{P}_\mathcal{R}(k)\sim 1.7 \times 10^{-6}$ 
for $k \sim 10^{2.2} \mathrm{Mpc}^{-1}$. For the smaller decay rate (longer lifetime), 
the limits are weaker.

\section{Conclusions}
If the dark matter are not annihilated, the decay model would be another important case. 
So, it is excepted that the UCMHs would contribute to the 
$\gamma$-ray flux due to the dark matter decay within them. 
On the other hand, the abundance of UCMHs is very important for modern cosmology. 
For example, it can be used to constrain the primordial 
curvature perturbations on small scales. In this work, we first investigated the 
integrated $\gamma$-ray flux from nearby UCMHs due to the dark matter decay. 
In the case of larger mass of UCMHs or smaller dark matter mass, e.g. 
$\mathrm{m_{\chi} = 200 GeV}$, $ \mathrm{M_{UCMHs}} = 10^{7} M_{\odot} $ 
and $b \bar b$ decay 
channel, the integrated $\gamma$-ray flux from the distance $d = 10$ kpc
would achieve the point source sensitivity of $Fermi$. 
Another important aspect is that we have studied the influence of different decay channels 
and density profiles of dark matter halo on the constraints 
of UCMHs abundance. One of the important parameters is the dark matter decay 
rate ($\Gamma$). Since the cosmological evolution, e.g. reionization, 
can be affected by the dark matter decay, this parameter can be constrained 
by the CMB observations. In this work, we have used the WMAP-7 years data 
to obtain the constraints and used these results to perform the following calculations. 
We considered two decay channels: $b \bar b$, $\tau^+ \tau^-$ 
and assumed each of the branching ratio $B_f = 1$. For the density profile 
of dark matter halo, we used NFW, Isothermal and Einasto models to 
obtain the final constraints. We found that the strongest constraints on the 
fraction of UCMHs comes from the smaller dark matter mass and the $b \bar b$ channel where 
$\mathrm{f_{UCMHs} \sim 5 \times 10^{5}}$ for $\mathrm{M_{UCMHs} \sim 10^{6} M_{\odot}}$. 
For all these three kinds of density profile, the strongest constraints are similar. 
The constraints on the fraction of UCMHs can be translated into the limits 
on the primordial curvature perturbations on smaller scales. We found that the 
strongest conservative limit is $\mathcal{P}_\mathcal{R}(k)\sim 1.7 \times 10^{-6}$ 
for $k \sim 10^{2.2} \mathrm{Mpc}^{-1}$ for the specific parameters of dark matter 
particles and the dark matter halo model. Although these results are weaker than the cases of
dark matter annihilation, they are still very useful if the dark matter 
particles are not annihilated.

\acknowledgments

Yang Yu-Peng thank Sun Weimin for improving the manuscript. 
We thank the referees for their very useful suggestions. 
Our MCMC chains computation was performed on the Shenteng 7000 system
of the Supercomputing Center of the Chinese Academy of Sciences.
This work is supported in part by the National Natural Science
Foundation of China (under Grant Nos 10935001 and
11075075).

\end{document}